\documentstyle[twoside,fleqn,espcrc2]{article}

\newcommand{\AmS}{{\protect\the\textfont2
  A\kern-.1667em\lower.5ex\hbox{M}\kern-.125emS}}

\title{Some Issues in Gauge Mediation}
\author{Michael Dine
\address{Physics Department, University
of California, Santa Cruz\\ 
  Santa Cruz, CA    95064}%
\thanks{Work supported by the U.S. Department
of Energy.}}
              
\begin{document}

\begin{abstract}
After briefly reviewing some arguments in favor of high
scale and low scale supersymmetry breaking, I discuss
a possible solution of the $\mu$-problem of gauge
mediated models.
\end{abstract}

\maketitle

\section{Introduction:  High Scale Vs. Low Scale Supersymmetry
Breaking}

In the absence of any experimental evidence
one way or the other,
a good case can be made for either high scale or low
scale breaking.  Perhaps the most compelling
argument in favor of high scale breaking is that such
breaking seems natural in
string theory.  For example, gaugino condensation
generates a superpotential
for moduli, 
\begin{equation}
W = e^{-S/M_p} M_p^3 \sim m_{3/2}M_p^2.
\end{equation}
with corresponding potential
\begin{equation}
V=e^K \left [ \vert {\partial W \over \partial
\phi} + {\partial K \over \partial \phi} W \vert^2
-3 \vert W \vert^2 \right ].
\end{equation}
If all of the moduli vary on the scale $M_p$, then
all terms here are of the same order.  ${\cal O}(1)$
effects in the Kahler potential can stabilize the moduli,
while the $-3\vert W \vert^2$ term
can cancel the cosmological constant\cite{bd}.

This is not the case if one has a low scale of supersymmetry
breaking.   The term $-3 \vert W \vert^2$ above
is then simply too small to cancel the cosmological
constant, unless there are some other interactions
with a much larger scale, introduced just
for this purpose.  One can shrug off this distinction,
saying that we don't understand the cosmological
constant.  After all, while for high scale breaking
all of the terms are of the same order, they still
must cancel to an extraordinary degree of precision.
Still, it is troubling that a new scale of interactions
seems to be required in the low scale case.

There are serious problems with the high scale
scenario as well.  One needs some mechanism to
understand squark and slepton degeneracy.
The various proposals to understand this are by now quite
familiar:  flavor symmetries,  luck, or some sort of
string miracle?

It is perhaps worth stating that if string theory
were to provide a theory of soft breakings (especially
prior to the discovery of supersymmetry), string
theory could find itself in a position similar
to that of weak interactions prior to the discovery
of the $W$ boson.  The Weinberg-Salam
theory was well established long before the
discovery of the $W$ and $Z$ because
it reproduced existing low energy data
and because it successfully predicted
the results of many new low energy
measurements.  Similarly,    string theory
could be convincingly confirmed if it provided
a theory of soft breakings, even though
experimental study of Planck scale physics
would be out of reach.

Low energy breaking has certain clear theoretical
advantages over high energy breaking\cite{lowenergy}:
\begin{itemize}
\item
It is more predictive.   One doesn't need to solve string
theory to determine the $106$ parameters of the MSSM.
\item
Flavor changing processes are automatically suppressed.
\item
The scale of supersymmetry breaking is perhaps not
too far removed from conceivable experiments.
\item
Interesting new phenomena are possible, such as
$\gamma$'s plus missing energy if the scale
is low enough\cite{ddrt}\cite{swy}.
\item
There exist real models which make predictions.
\end{itemize}

Apart from the question of scales mentioned above,
there are a number of troubling features of existing
approaches to model building.  Many of these are cosmological.
For example, in the framework of string theory, one expects
extremely light moduli.  This problem seems quite serious.
It might be solved if our vacuum lies at or near a point
of enhanced symmetry, where all of the moduli are charged
under some group.  It might also be solved by a period
of very late inflation\cite{riotto}.  Each of these ideas has difficulties
(but then so do  ideas for solving the more conventional
cosmological moduli problem).

More immediate are a set of problems on which there has been
some progress in the last year:
\begin{itemize}
\item  {\bf Complexity:}  The original models were quite
complex, with several layers of interaction.  During the
past year, there has been substantial improvement in this
picture, especially if one is willing
to give up the idea that the SUSY breaking
scale is ``nearby."  \cite{nelsonetal}
\item
{\bf  Fine Tuning:}  In minimal gauge mediation (MGM),
the negative correction to the Higgs mass from loops
containing the top squark is given by:
\begin{equation}
\delta m^2_{H_U} \approx -{12 y_t^2 \over
16 \pi^2} \ln(\alpha_s/\pi) \tilde m_t^2.
\end{equation}
Numerically, this is about 
$8$ times the positive contribution from
gauge loops, and about $70$ times the
contribution to the right handed sleptons.  But we
know that the masses of these latter particles
are greater than about $85$ GeV.  So obtaining
the correct breaking of $SU(2) \times U(1)$ requires
fine tuning.  This problem can perhaps be
ameliorated by relaxing some of the assumptions of
minimal gauge mediation\cite{agashe}\cite{dtw}.
For example, in the models
we discuss below, the messengers lie in a $10+ \overline{10}$
representation of $SU(5)$, and couple to several singlets.
The squark and slepton spectrum in such models is different
than in the case of minimal gauge mediation, and it is possible
for the singlet slepton masses to be close to those
of the electroweak doublets and color triplets.  
\item
{\bf Light Moduli:}  We have already alluded to the fact
that in the framework of string theory, low energy breaking
implies very light moduli.  These pose cosmological
challenges; they also
can be interesting,
in that they mediate long range forces\cite{dimopouloslongrange}.  
The cosmological problems might be solved if the minimum of
the moduli potential is at or very near a point
of enhanced symmetry.  Alternatively, it might be solved within
various inflationary scenarios.
\item
{\bf The $\mu$ problem}:  The $\mu$ problem has several aspects.
First, one can ask why $\mu$ isn't simply order $M_p$.
This might be a consequence of symmetries.    Alternatively,
in string theory, it has been known for a long time that $\mu$
may simply be small, at the classical level, more or less by
accident, and remain small as a consequence of
non-renormalization theorems.  $\mu$, however,
must not be zero but instead must be of order the weak
scale.  Similarly, $B_{\mu}$, the coefficient of
$H_UH_D$ in the {\it potential} should also be
of order the weak scale.
In the context of high scale breaking, this
can be arranged automatically.
This is not so easily arranged in the framework of gauge
mediation.  For example,
in models of gauge mediation there
is usually a singlet, $S$, whose scalar and $F$ components
have vev's, and which couples to the messenger fields.
One can simply suppose that this field couples
to the Higgs fields as well, with coupling $\lambda H_U H_D$.
One needs, however, that $\lambda S \sim 100 {\rm GeV}$.
But then $B_{\mu}$ is 
\begin{equation}
B_{\mu}= \lambda F_S = \lambda S{F_S \over S} \sim
10^7 GeV^2({S \over 30 {\rm TeV}})^2.
\end{equation}
Various solutions to this problem have been proposed\cite{pomarol,dnns}.
\end{itemize}

\section{Large $m^2_{H_D}$ and the $\mu$ Problem}

This last problem represents another
 rather severe fine tuning, and will be the focus
of the rest of this talk.  I want to consider here
another solution, inspired by ref. \cite{moreminimal}, which
has been developed with Scott Thomas and Jim Wells\cite{dtw}.
One of the features
of this work is a large value for $m^2_{H_D}$.
These authors made this proposal to solve a number of fine-tuning
difficulties.  Here, we exploit the fact that a large value
of $m^2_{H_D}$ can naturally solve
the $\mu$ problem in the form described above.  The main
point is quite simple, and can be seen by inspection of the
quadratic terms in the general Higgs potential:
\begin{equation}
V(H_U,H_D)= m^2_{H_D}\vert H_D \vert^2 - m_{H_U}^2
\vert H_U \vert^2
\end{equation}
$$~~~~~~~ + B_{\mu} H_U H_D.$$
If $m^2_{H_D}$ is large, then
\begin{equation}
\langle H_D \rangle = {B_{\mu} \langle H_U \rangle \over
m^2_{H_D}}.
\end{equation}
Note that if $m^2_{H_D} \gg B_{\mu}$, one
has large $\tan(\beta).$

How does one get large $m^2_{H_D}$?  One approach
is to give up the rigid philosophy of the
gauge mediated models, in which there is a complete
separation of the visible and messenger sectors.
For example, in the MGM, one has a singlet coupled to
a $5$ and $\bar 5$ of messengers:
\begin{equation}
W_{mgm}= S q \bar q + S \ell \bar \ell.
\end{equation}
In such a model, one can have, in principle, couplings
such as $H_D \ell \bar e$ and $H_D Q \bar q$.
Ordinarily, one assumes that these are forbidden by discrete
symmetries.  However, most Yukawa couplings in nature
are very small, for reasons we can only guess, and perhaps
these undesirable couplings are small as well.
If most of them are small enough (where $10^{-2}$ is
small enough) then one will still have no problem with
flavor changing neutral currents or with the unitarity
of the KM matrix.  For this talk, I will adopt this point
of view, and assume that, in addition to the
Yukawas above, only a few others are appreciable,
such as $H_D \ell \bar \tau$\cite{dns}.  In this case,
in addition to the
usual two loop gauge contributions familiar in gauge mediation,
there are additional {\it one-loop} contributions
to the masses of $H_D$ and $\bar \tau$.  These were
evaluated in \cite{dns}.  One obtains:
\begin{equation}
\delta m^2_{H_D} = -{b^2 \over 24}{1 \over 16 \pi^2} 
{\vert F \vert^4 \over M^6}
\end{equation}
where $b$ is the appropriate coupling constant from the
superpotential and $M$ is the mean mass of the multiplet.

So this model does produce a one loop mass for $H_D$,
but unfortunately of the wrong sign.  We have tried a variety
of approaches to obtaining a positive sign, and only
one seems to succeed without leading to other
undesirable consequences.
\begin{itemize}
\item
{\bf Larger $F_S$:}  If $F_S$ is large, then one cannot simply
expand in powers of $F_S$.  However, it turns out that for
all physically acceptable $F_S$, $m^2_{H_D}$ is negative.
\item
{\bf More $5+ \bar 5$'s:}    If there are no additional
singlets, this just leads to copies of the computation above.
\item
{\bf More singlets}:  Now one can obtain a positive contribution.
But there are other problems.  The most serious
is the appearance of a Fayet-Iliopoulos D-term for hypercharge.
In the MGM,
there is an accidental left-right symmetry of the messenger
sector, which forbids such a term.  This is not the
case in more general models.
\item{\bf $10,\overline{10} $ messengers:}  Now with one singlet,
one has the couplings
\begin{equation}
W_{mess}= \lambda_1 S Q \bar Q + \lambda_2 S U \bar U
\end{equation}
$$+ \lambda_3 S E \bar E + y H_D \bar Q U.$$
Now the required computation is different, but
one still finds a negative mass-squared, at order
$\vert F \vert^4 \over M^6$.  Again, this
persists to all orders in $F$.  But now one can add more singlets
to the model
without destroying the left-right symmetry of the model and
generating a one-loop Fayet-Iliopoulos term.
For a range of parameters, one finds that the masses are positive!
For example, if all of the supersymmetric masses are identical
and equal to $M$, and if one has a susy-breaking potential,
\begin{equation}
\delta V = \mu_Q^2 Q \bar Q + \mu_U^2 U \bar U,
\end{equation}
then one finds
\begin{equation}
m_{H_D}^2= {1 \over 2}{p^2 \over 16 \pi^2}
{(\mu_Q^2 - \mu_U^2) \over M^2},
\end{equation}
where $p$ is a combination of Yukawas in the superpotential.
So at least in this case, there is a positive one loop mass.
\end{itemize}

The model with the $10$ and $\overline{10}$, then,
provides at least an existance proof that one can
obtain a large mass for $H_D$.  It is interesting to take
the model seriously, and note a number of differences
with the MGM.  First, the gauge contributions to the low
energy soft breakings are characterized not by one parameter,
$\Lambda= F_S/S$, but three, which we can call
$\Lambda_Q$, $\Lambda_U$ and $\Lambda_E$.
In particular, if $\Lambda_Q > \Lambda_U,\Lambda_E$,
the ratio $\tilde m_t^2/\tilde m_{\ell}^2$ is reduced,
by a factor of $2/3$, improving the situation with regards
to fine tuning.  Even better is the situation where
$\Lambda_E \gg \Lambda_Q, \Lambda_U$.

\section{A model for $\mu$}

In light of these observations, let us reexamine the $\mu$-problem.
As we said in the introduction, we can generate $\mu$
through a coupling $\lambda_i S_i H_U H_D$,
where $S_i$ are the singlets which couple to the
messengers.  The couplings $\lambda_i$ must satisfy
\begin{equation}
\lambda_i \approx  ({\alpha_w \over 4 \pi} ),
\end{equation}
 if
$S \sim 100 {\rm TeV}$, in order that the $\mu$
term be of order the scale of electroweak symmetry
breaking.  This requires, if not an appreciable
fine tuning, a certain degree of good luck.  One of the
virtues of gauge mediation is that gaugino
masses are automatically one loop, while squark and
slepton masses-squared are automatically two loop
effects.  We can
try and go further in the present context and construct a model in
which the $\mu$ term is automatically a one loop effect,
while $B_{\mu}$ is automatically a two loop effect\cite{dtw}.

Suppose that none of the singlets, $S_i$, couples to $H_UH_D$,
but that there is another singlet, $S$, with couplings
\begin{equation}
W_S=\lambda S H_UH_D + S^3.
\end{equation}
Now at one loop there is a constribution to $m_S^2$ from loops
containing $H_D$.  This is similar to the usual negative contribution
to the Higgs from stop loops.  It automatically gives a negative
mass to $S$, which, since it is proportional to
$m^2_{H_D}$, is necessarily a two loop
effect.  So $\mu$ is automatically of one loop order, while
$B_{\mu}$ is necessarily a two loop effect.

While this is quite appealing, it is not immediately clear that
such a model can be realistic.  There are a number of issues
which must be faced:
\begin{itemize}
\item
{\bf $\tan(\beta)$}:  In this framework, $tan(\beta)$ is
necessarily large,
\begin{equation}
\tan(\beta) \approx {m_{H_D}^2 \over B_{\mu}}
\approx {16 \pi^2 \over ln(M^2/M_{H_D}^2) \times
{\rm couplings}}.
\end{equation}
There is a danger that $\tan(\beta)$ will be so large
that the bottom Yukawa coupling will become
non-perturbative.
\item
{\bf Other couplings}:  $\tan(\beta)$ and
other potentially problematic effects get larger
as the couplings $\lambda$ and $\lambda^{\prime}$
become smaller.  But these couplings are bounded
above by numbers of order $0.5$ if one requires that
they remain perturbative up to the unification scale.
\item
{\bf $m_{\tau}^2$}:  This mass receives a large negative
contribution from $H_D$ loops when $\tan(\beta)$
and $m^2_{H_D}$ are large.
\item
{\bf Light pseudoscalars}:  The model contains a relatively light
pseudoscalar (due to an approximate Peccei-Quinn symmetry,
broken by three loop effects).  The couplings of this
particle to the $Z$, however, are suppressed by powers
of $\tan(\beta)$, and this suppression seems adequate to
explain why the particle has not yet been observed.
\end{itemize}

A survey of the parameter space indicates that these
(and other) constraints can be satisfied\cite{dtw}.
For large $m^2_{H_D}$,
this is most easily achieved if the messenger scale is large;
this is perhaps appealing, given that many new models
prefer such a large scale.

\thanks{I wish to acknowledge the many insights of
my collaborators, S. Thomas and J. Wells.  I also
acknowledge the hospitality of the Stanford Center
for Theoretical Physics and of SLAC.}

\end{document}